\documentclass[twocolumn,aps,amsfonts,amsmath,prd,nofootinbib]
{revtex4-1}
\usepackage[T1]{fontenc}
 \usepackage[latin1]{inputenc}
\usepackage{amssymb}
\usepackage{amsmath}
\usepackage{enumerate}

\def\K{K{\"a}hler}
\def\be{\begin{equation}}
\def\ee{\end{equation}}
\def\ba{\begin{array}}
\def\ea{\end{array}}

\makeatletter

\usepackage{epsf}

\def\be{\begin{equation}}
\def\ee{\end{equation}}
\def\ba{\begin{eqnarray}}
\def\ea{\end{eqnarray}}
\def\beas{\begin{eqnarray*}}
\def\eeas{\end{eqnarray*}}
\def\sla{\raise.15ex\hbox{$/$}\kern-.57em}

\def\kk{{\cal K}}

\begin{document}

\title{\Large\bf General inflaton potentials  in supergravity}

\author{Renata Kallosh}
\author{Andrei Linde}
\author{Tomas Rube}

\affiliation{Department of Physics, Stanford University, Stanford, CA 94305, USA}


\begin{abstract}
We describe a way to construct supergravity models with an arbitrary inflaton potential $V(\phi)$ and show that all other scalar fields in this class of models can be stabilized at the inflationary trajectory by a proper choice of the   \K\, potential. 
\end{abstract}

\maketitle

\section{Introduction}

From a purely mathematical point of view, finding a realization of the chaotic inflation in the theory of a single scalar field $\phi$ is a nearly trivial exercise. One simply  finds a function $V(\phi)$ which is sufficiently flat in some interval of the values of the inflaton field  \cite{Linde:1983gd}. The simplest example of such potential is $m^{2}\phi^{2}/2$, or one can take any other function which approaches $\lambda_{n}\phi^{n}$ for $\phi > 1$ where $M_p=1$. If one wants to have inflation at $\phi < 1$, one can consider a function with a sufficiently flat maximum or an inflection point.

Finding a proper inflaton potential in supergravity is more difficult. 
The scalar potential in supergravity is a complicated function of the superpotential ${ W}$ and the \K\, potential. Usually the potential depends on several complex scalar fields. Therefore one has to investigate dynamical evolution in a multidimensional moduli space and verify stability of the inflationary trajectory. The main problem is related to the \K\, potential ${\cal K}$. The simplest  \K\, potential contains terms proportional to $\Phi\bar\Phi$. The F-term part of the potential is proportional to $e^{\cal K}\sim e^{|\Phi|^2}$ and is therefore much too steep for chaotic inflation at $\Phi \gg 1$. Moreover, the presence of the terms like $e^{|\Phi|^2}$ implies that the typical scalar masses are $\mathcal{O}(H)$, too large to support inflation even at $\Phi < 1$. 

From the point of view of an inflationary model builder, there is also an additional problem. One can always find $V(\phi)$ from ${W}$ and ${\cal K}$, but until the calculations are finished, one does not know exactly what kind of potential we are going to get. As a result, it is very difficult to solve the inverse problem: to find ${W}$ and ${\cal K}$ which would produce a desirable inflationary potential $V(\phi)$.

A partial solution to this problem was found in a recent paper  \cite{Kallosh:2010ug}. The main idea was to consider a theory of two scalar fields, $S$ and $\Phi$, with a \K\, potential which has a flat direction corresponding to ${\rm Re \, \Phi}$, and with a superpotential ${W} = S f(\Phi)$, where $f(\Phi)$ is an arbitrary holomorphic function. This class of models is a generalization of the simplest model of chaotic inflation in supergravity proposed long ago in \cite{Kawasaki:2000yn}. It was shown in  \cite{Kallosh:2010ug} that that for a certain class of functions $f(\Phi)$, the scalar field $\phi = \sqrt 2 {\rm Re}\, \Phi$ plays the role of the inflaton field.  Inflation occurs at $S = {\rm\, Im \Phi} = 0$, and this inflationary trajectory can be stabilized by a proper choice of the \K\, potential. In this class of models, one has a functional freedom of choice of the inflaton potential $V(\phi) = |f(\phi/\sqrt 2)|^{2}$. 

In this paper we will extend these results for a more general class of  \K\, potentials and explain how one can solve the inverse problem discussed above and obtain an arbitrary inflationary potential $V(\phi)$ using the  superpotential ${ W} = S f(\Phi)$. We will also show that one can always stabilize  the inflationary trajectory $S = {\rm Im \Phi} = 0$ by a proper choice of the curvature of the  \K\, manifold.

We will identify the field $S$ with the scalar component of the goldstino multiplet, and we will formulate the required conditions of stability of the inflationary trajectory in terms of the curvature of the \K\, geometry.

\section{General Inflationary Potential}\label{generalpot}

Consider a supergravity model with the superpotential
\be
W= Sf(\Phi)\, ,
\label{cond}
\ee
where $f(\Phi)$ is a real holomorphic function such that $\bar f(\Phi) = f(\Phi)$. Any function which can be represented by Taylor series with real coefficients has this property.

The real part of the field $\Phi$ will play the role of the inflation field. Meanwhile the fields $S$ and ${\rm Im}\, \Phi$ will be forced to vanish during inflation. As will be explained later, the scalar field $ S$ in this class of models belongs to the  goldstino supermultiplet; its fermionic partner is a goldstino.

One can impose certain simplifying conditions on the  \K\, potential ${\cal K}(\Phi,\bar\Phi,S,\bar S)$ without affecting the inflationary regime.
In what follows, we will assume that the \K\, potential ${\cal K}(\Phi,\bar\Phi,S,\bar S)$ is separately invariant with respect to the following transformations: 
\ba\label{S}
&& S\rightarrow  -S \ ,\\
&& \Phi\rightarrow \bar\Phi \  , \\
&&  \Phi\rightarrow\Phi+a,\ a\in\mathbb{R} \ .
\ea
The \K\ potential is invariant under the first transformation, in particular, if it is a function of $S^{2}+{\bar S}^{2}$ and $S\bar S$.
It implies that ${\cal K}_S= {\cal K}_{\bar S}=0$ at $S =0$ for all $\Phi$, where lower indices imply differentiation, e.g. ${\cal K}_{S \bar S} = \partial_{S}\partial_{{\bar S}}{\cal K}$. One then can show that the scalar potential $V(\Phi,\bar\Phi,S,\bar S)$ has an extremum with respect to $S$ and $\bar S$ at $S =0$, for all $\Phi$, i.e. $\partial_{S} V = \partial_{\bar S} V =0$ at $S = 0$. The potential of the field $\Phi$ at $S = 0$ is given by
\be
V(\Phi,\bar\Phi) = e^{{\cal K}(\Phi,\bar\Phi,0,0)}\ |f(\Phi)|^{2}\  {{\cal K}^{{-1}}_{S \bar S}}(\Phi,\bar\Phi,0,0), 
\ee
where we took into account that $W=0$  and $\partial_{\Phi} W=0 $ at $S=0$.
The matrix of the coefficients in front of the kinetic terms in the Lagrangian is block diagonal at $S =0$, i.e. ${\cal K}_{S\bar \Phi} = {\cal K}_{\bar S \Phi} =0$. 

The second assumption is satisfied, in particular, by any \K\, potential which behaves as some function of  $\Phi^{2}+{\bar \Phi}^{2}$ and $\Phi\bar \Phi$. Because $f(\Phi)$ is a real holomorphic function, the product $|f(\Phi)|^{2}$ also has this symmetry. Points with ${\rm Im}\, \Phi =0$ therefore correspond to extrema with respect to  ${\rm Im}\, \Phi$, i.e. the first derivative of $V(\Phi,\bar\Phi)$ with respect to ${\rm Im}\, \Phi$ vanishes at ${\rm Im}\, \Phi =0$. As we will see, under certain conditions to be discussed later, the second derivatives of the potential with respect to $S$ and $ {\rm Im}\, \Phi$ along the trajectory $S = {\rm Im}\, \Phi =0$ are large and positive, which means that it is stable with respect to the fluctuations of the fields $S$ and ${\rm Im}\, \Phi$. In this case, one can study inflation when the field ${\rm Re}\, \Phi$ plays the role of the inflaton field, and all other fields vanish.

Next, using (3) and (4), the \K\, potential can be written as a function of $(\Phi -\bar\Phi)^{2}$. Since ${\cal K}$ is flat in ${\rm Re}\, \Phi$, the exponential factor $e^{\cal K}$ is under control and the potential along the inflationary trajectory can be rewritten using the value of ${\cal K}$ at the origin:
\be
V = e^{{\cal K}(0,0,0,0)}\ |f({\rm Re}\, \Phi)|^{2}\  {{\cal K}^{{-1}}_{S \bar S}}(0,0,0,0).
\ee
Using \K\, invariance, one can always make a choice ${\cal K}(0,0,0,0) = 0$, which corresponds to rescaling of $f(\Phi)$. One can also take  ${{\cal K}_{S \bar S}}(0,0,0,0) = {{\cal K}_{\Phi \bar \Phi}}(0,0,0,0) =  1$. This corresponds to canonical normalization of the fields $S$ and $\Phi$, which can be achieved by rescaling of these fields at the point $S=\Phi =0$. Finally, since the \K\, potential does not depend on ${\rm Re}\, \Phi$, all fields remain canonically normalized along the inflationary trajectory $S = {\rm Im}\, \Phi =0$, for all values of ${\rm Re}\, \Phi$. The inflationary potential is therefore given by the amazingly simple and general expression
\be
V =f^{2}({\rm Re}\, \Phi)\ ,
\ee
where $f(\Phi)$ is an arbitrary real holomorphic function. 

Another way to present the result is in terms of the real canonically normalized fields $s$, $\alpha$, $\phi$ and $\beta$, related to the complex fields through
\be
S  ={1\over\sqrt 2}(s+i\alpha)\, , \qquad \Phi  ={1\over\sqrt 2}(\phi+i\beta) \ \ .
\label{cart}\ee
The potential of the inflaton field $\phi$ is then
\be
V(\phi) =f^{2}(\phi/\sqrt 2)\ \ .
\ee

There are two ways to use these results. First of all, one can give a long list of all possible potentials $V(\phi)$ which appear if one uses some simple real holomorphic function $f(\Phi)$, e.g. an arbitrary polynomial of $\Phi$ with real coefficients; see some examples in  \cite{Kallosh:2010ug}.

The second way to use these results is to find a supergravity theory that effectively behaves like single field inflation with an arbitrary potential $V(\phi)$ during slow roll inflation. 
This is done by defining a function $f(\phi/2)=\sqrt{V(\phi)}$, expanding it in powers of $\phi/\sqrt 2$, and analytically continuing this function. 
The part of the potential responsible for the inflationary regime is very flat and smooth and the expansion should give an accurate answer there.  Therefore replacing $\sqrt{V(\phi)}$ by a polynomial of a sufficiently high degree in $\phi/\sqrt 2$ and then performing the analytic continuation should reproduce the correct physics during inflation. 

In general, it may require more work to find an analytic continuation which accurately reproduces a given potential $V(\phi)$ both during inflation and after inflation. However, observations provide very limited information about $V(\phi)$ after inflation because it does not affect the large scale structure of the universe. Therefore one may find a family of many different functions $f(\Phi)$ which accurately reproduce the required shape of the potential $V(\phi)$ during inflation, and use only those of these functions which satisfy some general requirements on the post-inflationary shape of the inflationary potential. One of these requirements is that $f(\Phi)$ must almost exactly vanish at the minimum of the potential, to account for the smallness of the cosmological constant. Another requirement is the requirement of stability, which should be satisfied both during inflation and after it.

\section{Stability of inflationary trajectory}
\label{sec:stability}
Completing the investigation requires checking the stability of the inflationary trajectory $S = {\rm Im}\, \Phi =0$ with respect to small fluctuations of the fields $S$ and ${\rm Im}\, \Phi$. The stability conditions have a particularly simple form when  \K\ potential only depends on $S\bar S$ and this will assumed in what follows.\footnote{For more general  K\"ahler potentials where only (\ref{S}) is imposed $m_\beta$ is unchanged while the masses of $S$ are split:
$$m^{2}_{s\pm} = -\left(\, {\cal K}_{SS\bar S\bar S}\pm\left|{\cal K}_{SSS\bar S} - {\cal K}_{SS}\right|\, \right)\, f^{2} + ({\partial_{\Phi}f})^{2}. $$
Terms like  $S^2+{\bar S}^2$ in the K\"ahler potential thus make the  field $S$ less stable. Similarly, adding terms $\sim S^{3}$ to the superpotential does not improve the stability.} Because the \K\ potential now is a function of  $s^{2}+\alpha^{2}$, the mass of $s$ and $\alpha$ must be the same. Therefore, for investigation of stability it is sufficient to study $m^{2}_{s}= \partial^{2}_{s}V$ and $m^{2}_{\beta} = \partial^{2}_{\beta}V$ at the point $s = \alpha = \beta = 0$. After some algebra, one finds the masses of all fields orthogonal to the inflaton direction $\phi$
\ba\label{genstab}
m^2_{\beta} &=& 2 ( 1
- {\cal K}_{\Phi\bar\Phi S\bar S})\ f^{2} + ({\partial_{\Phi}f})^{2} - f\cdot  \partial^{2}_{\Phi}f \ ,\nonumber \\
m^{2}_{s} &= & m^{2}_{\alpha} = - {\cal K}_{SS\bar S\bar S}\, f^{2} + ({\partial_{\Phi}f})^{2} \ ,
\label{eqn:masses}
\ea
where all functions and their derivatives are calculated at the inflationary trajectory $S={\rm Im}\, \Phi =0$.
Furthermore, using 
\be
V=f^2\simeq 3H^2\ ,\quad \left(\frac{\partial_\Phi f}{f}\right)^2=\epsilon \ ,\quad \frac{\partial_\Phi^2 f}{f}=\eta-\epsilon \ ,
\ee
the masses can be rewritten as
\ba\label{genstab}
m^2_{\beta} &=&3 H^2 \left[ 2  (1
-  K_{\Phi\bar\Phi S\bar S})+2\epsilon -\eta\right]\ , \\
m^{2}_{s} &= & m^{2}_{\alpha}= 3 H^2\left[-K_{S\bar S S\bar S}+\epsilon\right] \ .
 \label{masss}\ea
 Absence of tachyonic instabilities along the inflationary trajectory 
\be
m^2_{\beta}, m^2_{s}, m^2_{\alpha} \geq 0
\ee 
 requires that
 \be 
 K_{\Phi\bar\Phi S\bar S}+{\eta\over 2} -\epsilon \,  \leq 1 \, ,\qquad 
  K_{S\bar S S\bar S}-\epsilon \, \leq 0 \ .
\label{KNoTachyon}
\ee

There is an interesting regime where  these masses are not tachyonic, but some of them are much smaller than $H$. Inflationary perturbations of these light fields will be produced, in addition to the usual perturbations of the inflaton field $\phi$. These perturbations may produce dangerous isocurvature perturbations of metric, but under certain conditions they produce usual adiabatic perturbations, as in the curvaton scenario  \cite{curva}.  This scenario is more complicated than the traditional single-field inflation scenario, but it has some interesting features, such as a possibility to obtain large non-gaussian perturbations of metric. To study this regime, one must keep all terms in equations (\ref{masss}), including the small slow-roll parameters. For example, one may design the $S$ field to play the role of the curvaton and require that  
 \be 
  0  < m^2_{s}   \ll  H^{2} \qquad \Rightarrow \qquad   -{1\over 3} \ll K_{S\bar S S\bar S}-\epsilon \, \leq 0 \ .
\label{Scurv}
\ee
Or, one may design the $\beta$ field to play the role of the curvaton and require that  
 \be 
  0  < m^2_{\beta}   \ll  H^{2}  \; \Rightarrow  \;   -{1\over 3} \ll 2  (1
-  K_{\Phi\bar\Phi S\bar S})+2\epsilon -\eta \, \leq 0 \ .
\label{bcurv}
\ee

 On the other hand, if we want to have a standard single-filed inflationary scenario where only the fluctuations of the inflaton field are generated, we must ensure that all other scalar fields are heavier than $H$:
\be
m^2_{\beta}, m^2_{s}, m^2_{\alpha}  \gtrsim H^{2} \ .
\label{l}\ee
During  slow roll single field inflation $\epsilon,\eta \ll 1$,  the last terms in the mass formula are often subdominant and can be dropped:
 \ba\label{genstab1}
m^2_{\beta} &\approx&6 H^2  (1
-  K_{\Phi\bar\Phi S\bar S})\ , \\
m^{2}_{s} &= & m^{2}_{\alpha}\approx -3 H^2 K_{S\bar S S\bar S}\ .
 \ea
In this case, the \K\ potential must obey the following conditions:
\be
 K_{\Phi\bar\Phi S\bar S} \lesssim {5\over 6} \, ,\qquad 
  K_{S\bar S S\bar S} \lesssim - {1\over 3} \ .
\label{K}\ee
It is easy to find \K\, potentials that satisfy these conditions, as will be demonstrated in the next section.

 In general, the behavior of the function $f$ may be constrained by the requirement of stability of the trajectory $S = {\rm Im \Phi} = 0$ {\it after} inflation, where the slow-roll parameters become large. However, as we already mentioned in the previous section, this would simply mean that one should modify the function $f(\Phi)$ after inflation. This does not affect the freedom of choice of $V(\phi)$ during inflation, which is the main quantity that we need to know to account for the observable consequences of inflation.

\section{Examples}

Let us consider some particular examples. We will start with a simple polynomial \K\, potential that is a generalization of the potential used in Ref. \cite{Kawasaki:2000yn}:
\begin{equation}
\mathcal{K} =  S \bar S - \frac{1}{2}(\Phi -\bar\Phi)^2 - {\zeta} (S\bar S)^2  +  \frac{\gamma}{2} S\bar S (\Phi -\bar\Phi)^2.
\label{K2}
\end{equation}
Note that the stabilizing terms $- {\zeta} (S\bar S)^2  + {\gamma\over 2} S\bar S (\Phi -\bar\Phi)^2$ was added to the \K\, potential of the model of \cite{Kawasaki:2000yn}. This \K\ geometry has $K_{\Phi\bar\Phi S\bar S}= -\gamma$ and $K_{S\bar S S\bar S}=-4\zeta$ and the stability conditions (\ref{K}) during inflation are, for any sufficiently flat $f(\Phi)$, $\gamma \gtrsim 5/6$ and $\zeta\gtrsim 1/12$. 

This can be verified for a simple model with $f(\Phi)=\lambda \Phi^n$, corresponding to $H^{2} = V/3 = \lambda^{2} \Phi^{2n}/3$. In this case (\ref{eqn:masses}) gives
\ba
m^2_{\beta} &=&   \lambda^{2} \Phi^{2n-2}\left(2(1+\gamma) \Phi^{2} + n\right)  \ ,
\nonumber\\
m^{2}_{s} &=& m^{2}_{\alpha}= \lambda^{2} \Phi^{2n-2}(4 \zeta   \Phi^{2} + n^{2}) \ .
\ea
Inflation in these models happen for $\Phi^{2} \gtrsim  {\rm max}\, \{n, n^{2}\}$, in which case
\ba
m^2_{\beta} &\approx&   6H^{2}\, (1+\gamma) \ ,
\nonumber\\
m^{2}_{s} &=& m^{2}_{\alpha} \approx  12 \zeta H^{2}\ .
\ea
 For $\gamma >-1, \zeta > 0$ both mass squared are positive and the inflationary trajectory is free from tachyonic instabilities even without the additional terms $- {\zeta} (S\bar S)^2  +  \frac{\gamma}{2} S\bar S (\Phi -\bar\Phi)^2$.  To avoid inflationary perturbations of these fields one should have $m_\beta^2,m_s^2,m^{2}_{\alpha}\gtrsim H^2$.  These fluctuations are suppressed for $\gamma\gtrsim - {5/6}$ and $\zeta\gtrsim1/12$.  Meanwhile, for $\zeta \ll 1/12$ one has $m_s^2 = m_{\alpha}^{2}\ll H^2$ during inflation.   In this case,  inflationary perturbations of the fields $s$ and $\alpha$ are generated, which may allow us to realize the curvaton scenario in this model \cite{DLM}. 
 
As a next example, consider the logarithmic \K\, potential which is used in the Jordan frame supergravity \cite{Einhorn:2009bh,Ferrara:2010yw,Lee:2010hj,Ferrara:2010in,Kallosh:2010ug}.
The  \K\, potential is
\ba
\mathcal{K} &=& -3\log \Bigl[ 1 + \frac{1}{6}(\Phi -\bar\Phi)^2 - \frac{1}{3}S \bar S +\zeta (S\bar S)^2/3  \nonumber\\ &-&   \frac{\gamma}{6} S\bar S (\Phi -\bar\Phi)^2\Bigr] .
\label{Ka}
\ea
In this case $K_{\Phi\bar\Phi S\bar S} = -\gamma+1/3$ and $K_{S\bar S S\bar S}=-4\zeta+2/3$ and the stability conditions with respect to the generation of inflationary perturbations of the fields orthogonal to the inflationary trajectory are $\gamma \gtrsim -1/2$ and $\zeta \gtrsim 1/4$.

Thus we found two different families of \K\, potentials which lead to stabilization of the inflationary trajectory, for any choice of the inflationary potential $V(\phi)$.

\section{goldstino and the geometric interpretation of stability}
The results above may be quite sufficient for cosmologists and for model builders. However, there are two issues which deserve additional discussion. First of all, when we studied stability of the inflationary trajectory, we used a particular representation $S  =(s+i\alpha)/\sqrt 2$,~ $\Phi  =(\phi+i\beta)/\sqrt 2$. However, one could represent these fields in many other ways, e.g. $S = s\, e^{i\theta}/\sqrt 2$,~ $\Phi = \phi\, e^{i\gamma}/\sqrt 2$. Is there any invariant way to study stability of the inflationary trajectory? Secondly, one may wonder what is the nature of the field $S$, which plays such an important role in the construction, but in the end vanishes. 

Moreover, one may want to understand the results above from a more general perspective of moduli stabilization in the near de Sitter geometry when one of the fields, the inflaton field, is light. Various aspects of generic supergravity and string theory moduli stabilization conditions in de Sitter vacua and during slow-roll inflation were studied before,  in particular in \cite{Denef:2004cf}, \cite{Covi:2008ea}. 

To answer these questions we take one step back and rederive the results above in the covariant formalism. The potential for ${\cal N}=1$ supergravity can be written as
\begin{align}
V &=e^\kk\left(g^{a\bar b}(\partial_a+\kk_a )W\, (\partial_{\bar b}+\kk_{\bar b})\bar W-3|W|^2\right)\cr
 &= \ g^{a\bar b}F_a\bar F_{\bar b}-3e^\kk |W|^2\ ,
\end{align}
where $F_a\equiv D_a e^{\kk/2} W =e^{\kk/2}(\partial_a W+\kk_a W$), $F^{\bar b}= g^{\bar b a} F_b$ and  $\bar F^{a}= g^{a \bar b } \bar F_{\bar b}$. The covariant derivative on a scalar is $D_a=\partial_a+w\, \kk_a$, where $w$ is the \K\ weight\footnote{The potential is invariant under the \K\ transformation 
$$\kk \rightarrow\kk+ f+\bar f\, ,\,\, W\rightarrow W e^{-f}.
$$
An object that transforms as $A\rightarrow A e^{-(f- \bar f)/2}$ has weight $w$. For example, $e^{\kk/2}W$ has $w=1/2$ and $V$ has $w=0$. The anti-holomorphic derivative is $D_{\bar a}=\partial_{\bar a}-w\, \kk_{\bar a}$.}, and appropriate metric connections $\Gamma_{ac}^b$($\Gamma_{\bar a\bar c}^{\bar b}$) are added when acting on tensors. For canonically normalized fields the stability is determined by the eigenvalues of the mass tensor
\be\label{eqn:masstensor}
{\cal M}= \left(\begin{array}{cc} \bar D_{\bar a}\partial_b V & \bar D_{\bar a}\partial_{\bar b} V \cr 
 D_{ a}\partial_{ b} V  &  D_{ a}\partial_{\bar b} V \end{array}\right)\ .
\ee
The holomorphic-anti-holomorphic part is
 \ba\label{eqn:holomorphic-antiholomirphic1}
{\cal M}_{a\bar b}&= &(g_{ a \bar  b} g_{c \bar d}- g_{a \bar d} g_{b \bar c} -R_{a\bar b c \bar d}) \bar F^c  F^{\bar d}  \nonumber\\
&-& 2g_{a\bar b} e^{\cal K} |W|^2 + D_a F_c \bar D_{\bar b} \bar F^c \ .
\label{massHAH}\ea
Supersymmetry is spontaneously broken when $F_a\neq 0$ and it defines the goldstino direction in the field space $U_a\equiv F_a/\sqrt{F_b F^b}$, $U_aU^a=1$. This can be seen from the goldstino fermion $\eta= F_a \chi^a$ that is eaten by the gravitino.  A special role of the goldstino direction in the moduli space was explained  in \cite{Covi:2008ea}.

The class of models considered in this paper have, during inflation, $W=0$ and spontaneously broken supersymmery $F_S=f(\Phi)\ne 0$ and $F_\Phi=0$.  This illuminates the role of $S$; because the goldstino is $\eta=F_a\chi^a=f(\Phi)\chi^S$,   $S$ belongs to the goldstino supermultiplet as it is a superpartner of the goldstino fermion. Furthermore, it is the non-zero $F_S$ that generates the inflaton potential:
\be
V=  F_S \bar F^S= 3H^2 \ .
\ee
We now simplify (\ref{eqn:masstensor}) along the inflaton trajectory in the limit where the slow roll parameters $\epsilon$ and $\eta$ are negligeble. Then the term  $D_a F_c \bar D_{\bar b} \bar F^c$ in (\ref{eqn:holomorphic-antiholomirphic1}) can be dropped and the holomorphic-antiholomorphic part of the mass formula simplifies
 \be
{\cal M}_{a\bar b}=(g_{ a \bar  b} g_{c \bar d}- g_{a \bar d} g_{ \bar b c} -R_{a\bar b c \bar d}) \bar F^c  F^{\bar d} \ .
\label{hah} \ee
A couple of points are to be made here. First, because the \K\, potential is of the form $\kk(S\bar S,(\Phi-\bar\Phi)^2)$ all odd derivatives vanish along the inflaton path and $\Gamma^b_{ac}=\kk_a=0$. The Riemann curvature simplifies to 
\be
R_{a \bar b c \bar d}= {\cal K}_{a  c \bar b \bar d} - \Gamma^e_{ab} g_{e\bar e} \Gamma^{ \bar e}_{\bar c\bar d}={\cal K}_{a  c \bar b \bar d}.
\ee
Thus both $R_{S\bar S S\bar\Phi}$ and the off-diagonal ${\cal M}_{S\bar\Phi}$ vanish. Second, using $U_a$ a sectional Ricci tensor in the goldstino direction can be defined, namely
 \be
R_{a\bar b}^{F_S}\equiv R_{a\bar b c\bar d} \bar U^c U^{\bar d}=  R_{a\bar b c\bar d} {\bar F^c  F^{ d}\over  F_b \bar F^{ b}}=K_{a\bar b S\bar S}\ .
 \label{curv}\ee
This suggests that that the constraints on $\kk_{S\bar S S\bar S}$ and $\kk_{\Phi\bar\Phi S\bar S}$ in 
 (\ref{KNoTachyon}) and (\ref{K})  should be interpreted as constraints on the sectional curvature tensor. In terms of the sectional curvature we have 
\be
 {\cal M}_{a\bar b}=(g_{ a \bar  b}  - U_a \bar U_{\bar b}  -R_{a\bar b}^F )V
 \label{eqn:MassSimple} \ .
 \ee
For $S$ and $\Phi$ this simplifies to 
 \ba
{\cal M}_{\Phi \bar \Phi}&=&(g_{ \Phi \bar  \Phi}  -R_{\Phi \bar \Phi} ^{F_S} )V \ ,\nonumber\\
{\cal M}_{S \bar S}&=& -R_{S \bar S}^{F_S}  V \ ,
 \ea
where the first two terms in the $S\bar S$ mass formula (\ref{eqn:MassSimple}) canceled.

Finally, using the sectional curvature a necessary condition for the absence of the tachyons in de-Sitter vacua can be stated as a geometric constraint on the \K\, manifold projected into a  goldstino direction:
\be
R_{a\bar b}^F  \leq g_{a\bar b}-\frac{F_a\bar F_{\bar b}}{V}.
 \label{simple}
\ee
 We now turn to the off-diagonal block ${\cal M}_{a b}$. To lowest order in slow roll parameters $f(\Phi)$ is a constant and the potential can be written in terms $V(S\bar S,(\Phi-\bar\Phi)^2)$. The $U(1)$ invariance in $S$ implies ${\cal M}_{SS}=0$ and the mass matrix for $S$ is
\be
\left(\begin{array}{cc} {\cal M}_{S\bar S}  &  0 \cr 
0  &  \, \, {\cal M}_{S\bar S}  \end{array}\right)  .
\ee
Because $V$, for small slow roll parameters, is an approximate function of $(\Phi-\bar\Phi)^2$ the derivatives are antisymmetric in $\Phi$ and $\bar\Phi$:
\be
D_\Phi V=- D_{\bar \Phi}V\ , \qquad   {\cal M}_{\Phi \Phi}=-{\cal M}_{\Phi\bar \Phi}  \ ,
\label{shift}\ee
and the $\Phi$ mass matrix is
\be
 \left(\begin{array}{cc} {\cal M}_{\Phi\bar \Phi}  &  -{\cal M}_{\Phi\bar \Phi}  \cr 
-{\cal M}_{\Phi\bar \Phi}   &  \, \, {\cal M}_{\Phi\bar \Phi}  \end{array}\right)\  \quad  \Rightarrow \quad 
\left(\begin{array}{cc} 2{\cal M}_{\Phi\bar \Phi}  &  0 \cr 
0  &  \, \, 0 \end{array}\right),
\ee
where the arrow indicates diagonalization. Here the inflaton $\phi$ is the nearly  massless state while $\beta$ has mass $2{\cal M}_{\Phi\bar\Phi}$. Using (\ref{eqn:MassSimple}) we now have a geometric interpretation of the results of section \ref{sec:stability}:  $\alpha$, $s$ and $\beta$ are all stable with suppressed inflationary perturbations for sectional curvatures satisfying (cf. (\ref{K}))
\be
R_{\Phi \bar \Phi }^{F_S} \lesssim -{5\over 6} \ , \qquad 
 R_{S\bar S }^{F_S} \lesssim  -{1\over 3}  \ .
\label{SC}\ee
which provides that $m^2_{\beta}, m^2_{s}, m^2_{\alpha}  \gtrsim H^{2}$.

However, as already mentioned, one may also be interested in a regime where the supergravity fields orthogonal to the inflationary trajectory are stable, but light, with potentially big inflationary fluctuations \cite{curva,DLM}. In such case one would like to have the masses of these fields significantly smaller than the Hubble scale. For example, if one would like to have the $S$ field light, one would require that $0  < m^2_{s}   \ll  H^{2}$.
 The restriction on the \K\, manifold curvature in such case is (cf. (\ref{Scurv}))
:
\be
- {1\over 3} \ll    \, R_{S\bar S }^{F_S}  -\epsilon < 0  \ .
\label{cs}\ee
If one would like to have the field $\beta$ light, $0  < m^2_{\beta}   \ll H^{2}$,
the  \K\, manifold curvature has to be restricted (cf. (\ref{bcurv}))
\be
 - {1 \over 3} \ll   \, 2 (R_{\Phi \bar \Phi }^{F_S} -1)-2\epsilon +\eta    < 0 \, . 
\label{cb}\ee
Thus, we derived the geometric stability conditions on inflationary trajectory $S = {\rm Im \Phi} = 0$. These conditions  involve the curvature of the \K\, manifold projected into the plane defined by the goldstino direction $S$ in the moduli space.  During inflation, supersymmetry is broken spontaneously in this direction, which is 
orthogonal to the inflaton. The scalar partner of a goldstino vanishes during inflation, which leads to a significant simplification of the analysis of the potential and of the stability of the inflationary trajectory.

\section{Conclusions}
In this paper, we identified a class of supergravity models where one can obtain an arbitrary inflationary potential $V(\phi)$.   All scalar fields along the inflationary trajectory have canonical kinetic terms and minimal coupling to gravity. One may wonder whether the stability conditions for the inflationary trajectory could lead to some constraints on the choice of the inflationary potential.
 
Fortunately, in this class of models  one can make functional adjustments to the \K\, potential and stabilize the inflationary trajectory without changing $V(\phi)$. Here we studied few  types of stability conditions. 

To achieve a single field inflation we require that all scalar fields but inflaton should have masses greater than $H^{2}$. This is necessary to avoid a tachyonic instability, and, in addition, to avoid generation of long-wavelength perturbations of all fields except the standard inflaton perturbations. We have found that these stability conditions  do not put any constraints on the choice of the potential  $V(\phi)$ during slow-roll inflation. They can be satisfied by the choice of the \K\,  geometry, see  (\ref{K}), (\ref{SC}). Thus, from a purely mathematical standpoint, one could use full functional freedom to tune the inflaton potential in supergravity, if it is required by observations.

We also studied the possibility that not only the inflaton field $\phi$, but also some other scalar fields have positive mass squared which is much smaller than $H^{2}$.  As shown in (\ref{Scurv}), (\ref{cs}) and (\ref{bcurv}), (\ref{cb}), this can be achieved by tuning the \K\, potential. This allows one to use the curvaton mechanism and add a controllable amount of non-gaussian adiabatic perturbations generated by fluctuations of the light scalar fields orthogonal to the inflationary trajectory.

We are grateful to  M. Hertzberg, S. Kachru, A. Marrani, V. Mukhanov, M. Noorbala,  L. Senatore, E. Silverstein and  A. Westphal   for  enlightening discussions.   This work was supported by NSF grant PHY-0756174. TR is a William R. and Sara Hart Kimball Stanford Graduate Fellow.

\end{document}